\documentclass[aps,pre,onecolumn,showpacs,groupedaddress,amsmath,amssymb]{revtex4}
\usepackage{graphicx}   
\usepackage{dcolumn}    
\usepackage{bm}         

\begin{document}

\preprint{Manuscript for submitting to Physical Review E}

\title{Influence of Temperature on Neuronal Excitability in Cochlear Nucleus}

\author{Ting Zeng$^{1}$}
\author{Jiafu Wang$^{1,2}$}
    \email[]{Correspondence:  jasper@whut.edu.cn}
\author{Shenbing Kuang$^{1}$}

\affiliation{ 1 School of Sciences, Wuhan University of Technology, Wuhan 430070, China\\
2 State Key Laboratory of Advanced Technology for Materials Synthesis and
Processing, Wuhan 430070, China}

\date{\today}

\begin{abstract}
The influence of temperature on neuronal excitability is studied by
numerical simulations on the spiking threshold characteristics of
bushy cells in cochlear nucleus periodically stimulated by synaptic
currents. The results reveal that there is a cut-off frequency for
the spiking of bushy cell in a specific temperature environment,
corresponding to the existence of a critical temperature for the
neuron to respond with real spikes to the synaptic stimulus of a
given frequency, due to the finiteness of spike width. An optimal
temperature range for neuronal spiking is also found for a specific
stimulus frequency, and the temperature range span decreases with
increasing stimulus frequency. These findings imply that there is a
physiological temperature range which is beneficial for the
information processing in auditory system.
\end{abstract}

\pacs{\ 87.\,19.\,Dd, 87.\,10.\,+e, \ 87.\,17.\,Aa, 87.\,19.\,La}   

\maketitle

Electric excitability has been attracting much attention for decades, for it is
the basis of information processing and is essential to coding in neural
systems. The excitability of a neuron originates from integrated effect of the
kinetics \cite{HH1952}, as well as the spatial distribution \cite{Segev2000},
of ion channels on the cellular membrane in proper environment, and determines
the functional properties when responding to external stimulus, such as the
frequency sensitivity \cite{LWW1999,Liu2001} or selectivity \cite{Kuang2006},
interaural time difference (ITD) sensitivity in auditory system
\cite{Lohuis2000}, etc. Excitability and response properties of a neuron may
vary in different environmental conditions of, say, temperature and pH values
\cite{Schweitzer2000}. Recently, the influence of temperature on the
excitability of rat suprachiasmatic nucleus neurons has been investigated
experimentally, and the results reveal that there is a temperature-sensitive
range for the neuronal activities; this may provide cues to the circadian
synchronized rhythmicity \cite{Burgoon2001,Miller1994}. Biophysically,
temperature may influence the functioning of a neuron through the temperature
dependence of various ion channel conductances and time constants of channel
activation/inactivation variables \cite{Rothman2003}; hence changing
temperature alters the basic properties of excitable neuron, such as the
membrane potential, the input resistance, the shape and amplitude of action
potentials, and the propagation of spikes
\cite{Griffin1995,Cabanes2003,Cao2005,Maxim2000}. Up to date, however, there
has been little theoretical investigations of the influence of temperature on
neuronal excitability in literature.

Neuronal excitability can be described by the firing properties, like the
spiking threshold of the neuron responding to periodic stimulus
\cite{Kuang2006}. Characteristics of the spiking threshold of excitable neurons
have been discussed in several studies
\cite{LWW1999,Kuang2006,Wang1997,Yu2001,Xie2004}, where the stimuli applied are
mostly sinusoidal. More realistically, in fact, the stimulus to a post-synaptic
neuron is often described by a current with alpha-function channel conductance
\cite{Rall1967}, which is used in the present study. Comparable with
Ref.\,\cite{Kuang2006}, this work focuses on the effect of temperature on the
spiking threshold of an auditory neuron periodically stimulated by excitatory
post-synaptic current (EPSC). The frequency dependence of the spiking threshold
varies in different temperatures. Our results on the temperature dependence of
the spiking threshold reveal that, for the EPSC stimulus with a given
frequency, there is a temperature sensitive range for the neuron, and that this
range reduces as the stimulus frequency increases. These imply that there is a
physiological temperature range which is beneficial to the information
processing in auditory system.

The model used in this work is presented by Rothman and Manis (RM)
\cite{Rothman2003} for bushy cells in ventral cochlear nucleus of auditory
midbrain based on electrophysiological experiments \cite{Rothman2003e}. It
consists of a single electrical compartment with a membrane capacitance ($C$)
connected in parallel with a fast-activating slow-inactivating low-threshold
K$^{+}$ current ($I_{LT}$), a high-threshold K$^{+}$ current ($I_{HT}$), a
fast-inactivating TTX-sensitive Na$^{+}$ current ($I_{Na}$), a
hyperpolarization-activated cation current ($I_h$), a leakage current
($I_{lk}$), and an excitatory synaptic current ($I_E$). The membrane potential
$V$ is described by the following first-order differential equation:
\begin{equation*}
\begin{array}{rcl}
C\frac{dV}{dt} &=\;& G_{Na}m^{3} h (V_{Na}-V) + G_{LT}w^{4} z (V_K-V) \\
               &  & + G_{HT}[\varphi{}n^{2}+(1-\varphi)p](V_K-V)     \\
               &  & + G_hr(V_h-V) + G_{lk}(V_{lk}-V) + I_E,
\end{array}
\end{equation*}
where $G$'s denote the maximum channel conductances, and $V_{Na}$, $V_K$,
$V_h$, and $V_{lk}$ are the reversal potentials for potassium, sodium, cation,
and leakage channel currents, respectively. The channel currents are governed
by some activation/inactivation variables $x$ ($x = w, z, n, p, m, h, r$)
satisfying the differential equation $dx/dt=[x_\infty(V)-x]/\tau_x(V)$, where
$\tau_x$ and $x_\infty$ are the voltage-dependent time constant and the
steady-state value of $x$, respectively. All the model parameters used here are
the same as in Ref.\,\cite{Rothman2003}. In the present study, the EPSC is
modeled by $I_E=g_E(t)(V_E-V)$, with $V_E$ being the reversal potential of the
EPSC (usually chosen as $V_E = 0$ mV for excitatory synapse) and $g_E(t)$ being
the time-dependent post-synaptic conductance in response to a sequence of
synaptic stimuli turn on at different time $t_i$, $g_E(t) = \sum_{i} G_{syn}
\alpha(t-t_i)$, where $G_{syn}$ determines the peak of synaptic conductance and
$\alpha(t) = (t/\tau_E)\exp[1-(t/\tau_E)]$, $t > 0$, with $\tau_E$ determining
the time to reach the stimulus peak (chosen as $\tau_E = 0.2$ ms in this
study). Standard fourth-order Runge-Kutta algorithm is applied to solve the
differential equations. Noticeably, in comparison to the definition in
Ref.\,\cite{Kuang2006}, the spiking threshold here is characterized as the
critical value of $G_{syn}$ for a neuron to fire. The implementation of
calculating the spiking threshold is based, still as usual
\cite{LWW1999,Kuang2006,Wang1997,Yu2001}, on estimating whether the membrane
potential $V$ of the stimulated neuron can exceed a voltage threshold $V_{th}$
(chosen as $V_{th} = -25$ mV here).

The parameters given in the RM model \cite{Rothman2003} are obtained from the
experiments in vitro \cite{Rothman2003e} operated at room temperature $T =
22\,^{\circ}$C. In order to include the influence of temperature on the
excitability of bushy cells, we will take the suggestion in
Ref.\,\cite{Rothman2003} that all the model time constants $\tau_x$ are divided
by a Q10 factor of 3 while all the maximum channel conductances (except for
$G_{syn}$) are multiplied a Q10 factor of 2. These Q10 values are
approximations to those reported for Na$^+$ currents \cite{Na_Q10} and K$^+$
currents \cite{HH1952,Kros1990}.

Our investigation begins with the frequency characteristics of the spiking
threshold of the RM model neuron stimulated by periodic EPSCs (with frequency
$f_s$) at $T = 22\,^{\circ}$C, the temperature at which most of the
experimental data in vitro was recorded in the model construction
\cite{Rothman2003,Rothman2003e}. The frequency dependence of the spiking
threshold is illustrated in Fig.\,1. Different from the case of sinusoidal
stimuli (see the results in Ref.\,\cite{Kuang2006}), the spiking threshold in
the case of more realistic synaptic stimulus here depends slightly on the
stimulus frequency in low frequency range. This may result from the effect of
the refractory period of the neuron; when the stimulus frequency period is long
enough for the membrane potential to return to the resting value, the spiking
threshold will change little with respect to the stimulus frequency. The
existence of a minimum spiking threshold at the frequency about 40Hz (see
inset\,(a) in Fig.\,1) is consistent with the previously reported frequency
sensitivity phenomenon in bushy cells \cite{Kuang2006}, as well as in
Hodgkin-Huxley (HH) \cite{Liu2001,Yu2001} and Hindmarsh-Rose
\cite{Liu2001,Wang1997,Yu2001} neurons, likely due to the intrinsic oscillation
in the excitable neuronal system \cite{Kuang2006,Wang1997}.

Furthermore, the frequency dependence of spiking threshold behaves an abrupt
jump in the high frequency range. As a matter of fact, when the stimulus
frequency is higher than the frequency range of the abrupt jump, the neuron
does not really fire any spike no matter how large the stimulus amplitude is,
though the membrane potential may exceed $V_{th}$ (see an example in the
inset\,(b) in Fig.\,1, where one may find that the neuron does not experience
any refractory period, nor even hyperpolarization). If one uses
three-compartment model in simulation, one will find that such kind of
non-spiking response can not propagate to the end of a long axon. Therefore, we
would refer to the frequency of the abrupt jump as the cut-off frequency for
the neuron to fire spike. The existence of this cut-off frequency may come from
that the neuron needs a time width to complete a spike; if the stimulus period
is shorter than a complete spike width, the neuron will not fire any real
spike.

The frequency characteristic of the spiking threshold may vary at different
temperatures, for temperature changes the maximum channel conductances and the
time constants of the channel activation/inactivation variables. Some examples
for bushy cell at $T = 10\,^{\circ}$C, 30$\,^{\circ}$C, and 38$\,^{\circ}$C are
illustrated in Fig.\,2, where one can see that, at different temperatures, the
frequency characteristics of spiking threshold are very similar, but the values
of the cut-off frequencies are distinct; the higher is the temperature, the
higher is the cut-off frequency. For each temperature the cut-off frequency is
close to the reciprocal of the corresponding spike width, confirming that the
existence of cut-off frequency for bushy cell is due to the finiteness of spike
width of the neuron. Since increasing temperature shortens the time constants
of channel variables and makes the neuron respond faster to external stimulus,
the spike width decreases and correspondingly the cut-off frequency increases
(see Fig.\,2 and inset therein).

Figs.\,1 and 2 show that at a given environmental temperature there is a
reliable frequency range of stimuli for a bushy cell to respond to fire real
spikes (propagable along axon). This will be very important to neuronal
communication and/or coding. If the frequency of an external stimulus (such as
sound signal) is too high, the neuron will not respond properly, nor will
process the information carrying by high frequency signals. Seeing that a bushy
cell can not work well for very high frequency stimuli at room or body
temperature, one may draw the conclusion that the bushy cells in auditory
midbrain will not be responsible for perceptual task of frequency
discrimination/coding in auditory information processing. This conclusion is
consistent with the results in Ref.\,\cite{Kuang2006}.

For a neuron stimulated by EPSCs of a given frequency, on the other
hand, how does the neuron's excitability depend on the environmental
temperature? The temperature dependence of spiking threshold of a
bushy cell stimulated by periodic EPSCs of $f_s = 100$\,Hz,
as an example, is illustrated in Fig.\,3. One can see that the spiking
threshold exhibits a global minimum in an environmental temperature range where
the bushy cell needs weakest synaptic stimulus to initiate spikes, indicating
the occurrence of optimal use of synaptic transmission. This result implies
that there is a sensitive temperature range for neuronal
activitities/communication of bushy cells. To further explore the mechanism
underlying the emergence of sensitive temperature range, we examine the effects
of temperature regulation through the maximum ion conductances (inset (a) in
Fig.\,3) and through the time constants of channel gating variables (inset (b)
in Fig.\,3), respectively. If only the influence of temperature on the maximum
ion channel conductances is considered, the spiking threshold has a monotonous
temperature dependence; in contrast, the sole temperature effect via channel
activation/inactivation rates yields a temperature dependence of spiking
threshold similar to the control result (Fig.\,3). Thus we conclude that the
emergence of the temperature sensitive range results mainly from the effect of
temperature upon ion channel kinetics. Theoretically, a U-shaped dependence of
a physical quantity characteristic may results from the competition of the
impacts of (at least) two dominant factors/aspects. The temperature dependence
of spiking threshold (as shown in Figs.\,3 and 4) comes likely from the
competition of the effects of temperature regulation on channel kinetic rates
of activation and of inactivation variables (cf. Ref.\,\cite{Fitzhugh1966} for
HH model). Detailed study will appear elsewhere.

In Fig.\,3  one may also find a sudden swerve (at about
6\,$^{\circ}$C) on the temperature-dependent spiking threshold
curve. Actually when temperature is lower than the temperature at
the swerve, the bushy cell does not fire any real spike; this is
consistent with the existence of cut-off frequency described above
(cf. Fig.\,2). Such kind of critical temperature for the synaptic
stimulus of a given frequency, below which the neuron does not work
well, also exists in the temperature dependence of spiking threshold
subject to periodic EPSCs of other frequencies (see examples in
Fig.\,4). Fig.\,4 also shows that there exists a specific
temperature-sensitive range for each stimulus frequency, while the
temperature range span decreases with increasing stimulus frequency.

In summary, the influence of temperature on neuronal excitability has been
investigated by numerically simulating the characteristics of spiking threshold
of bushy cell stimulated by periodic EPSCs of different frequencies in various
environmental temperatures. We find that, at a given environmental temperature,
there is a cut-off frequency for the spiking of bushy cell, leading to an open
question how the signal of high frequency is processed in auditory system. On
the other hand, for the synaptic stimulus of a given frequency, there exists an
optimal temperature range for the excitable neuron to respond most sensitively
to external input; the essence of this temperature dependence for neuronal
excitability may account for the temperature-sensitive properties found in
other kind of neurons \cite{Burgoon2001}. Furthermore, the existence of a
critical temperature for neuronal spiking found in this study may relate to the
mechanism of hibernator activities in winter.

This work is supported by the Key Project of Chinese Ministry of Education
(Grant No. 106115).

\newpage
\begin{figure}
   \includegraphics{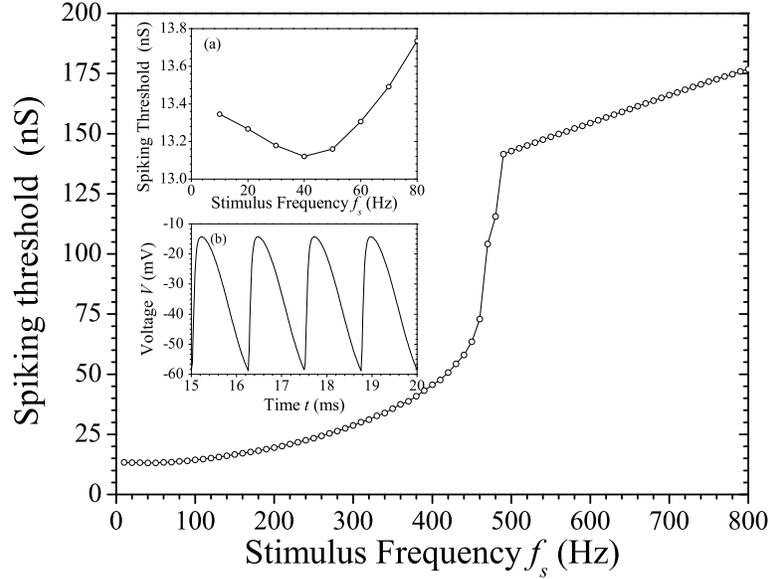}
\caption{\label{Fig1} The frequency dependence of spiking threshold of a bushy
cell stimulated by periodic EPSCs at $T = 22\,^{\circ}$C. Inserts: (a) Details
for the frequency range around the spiking threshold minimum. (b) The response
of the bushy cell to periodic EPSCs with $G_{syn} = 500$\,nS (larger than the
spiking threshold) at $f_s = 800$\,Hz (higher than the cut-off frequency). }
\end{figure}

\newpage
\begin{figure}
   \includegraphics{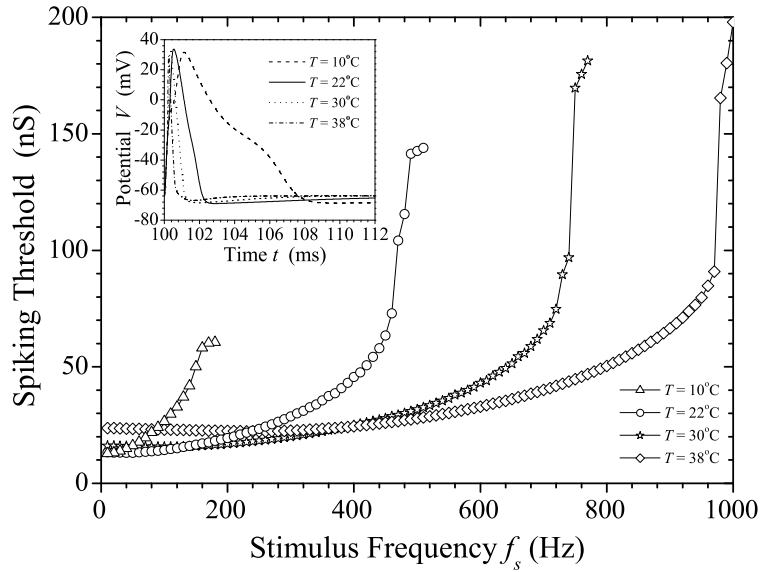}
\caption{\label{Fig2} Frequency dependence of spiking threshold of bushy cell
stimulated by periodic EPSCs at different temperatures. (For each temperature
only the reliable frequency range is shown.) Inset: The responses of a bushy
cell to a synaptic stimulus (with $G_{syn} = 100$\,nS), showing spike widths,
at different temperatures. }
\end{figure}

\newpage
\begin{figure}
   \includegraphics{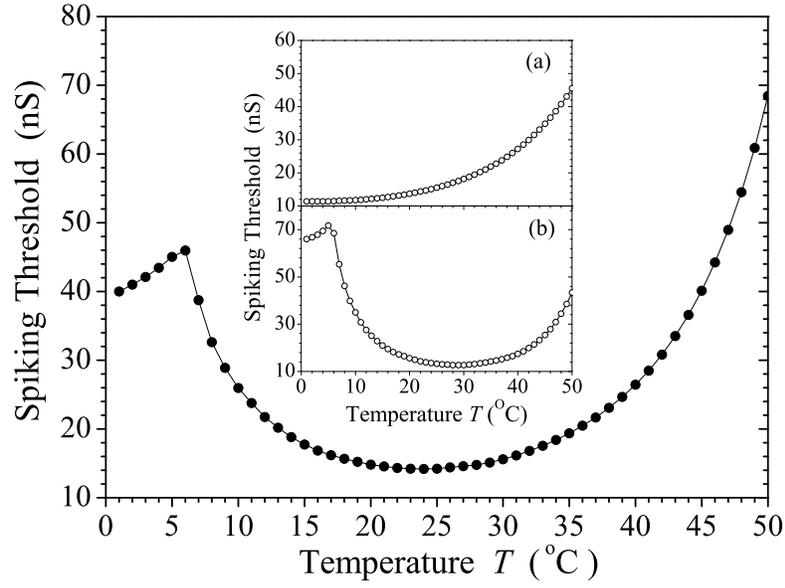}
\caption{\label{Fig3} The temperature dependence of spiking threshold of a
bushy cell stimulated by periodic EPSCs of $f_s = 100$\,Hz. Main: Both the
channel conductances and the time constants of channel variables are regulated
by the Q10 factors of temperature proposed in Ref.\,\cite{Rothman2003}. Insets:
(a) Only the temperature regulation on the channel conductances is considered.
(b) Only the temperature regulation on the time constants of channel variables
is considered. }
\end{figure}

\newpage
\begin{figure}
   \includegraphics{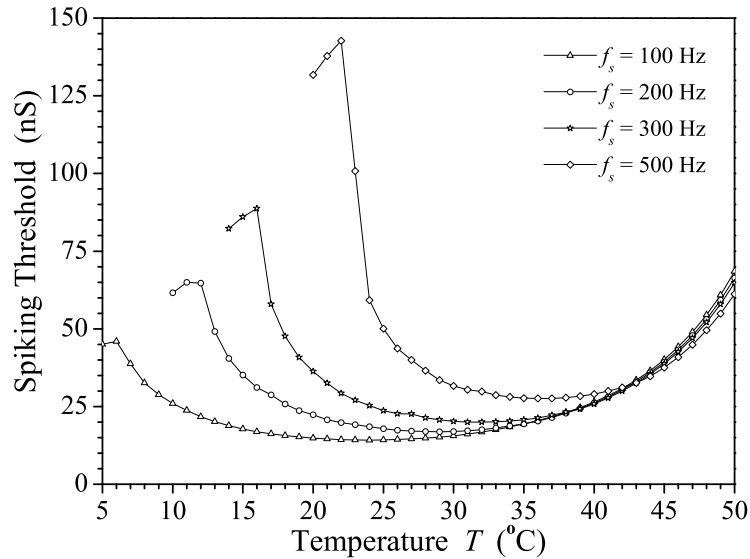}
\caption{\label{Fig4}
 The temperature characteristics of spiking
threshold of bushy cell for the stimuli of different frequencies. (For each
stimulus frequency only the reliable temperature range is shown.) }
\end{figure}

\end{document}